# MODELS AND EXPERIMENTAL RESULTS FROM THE WIDE APERTURE NB-TI MAGNETS FOR THE LHC UPGRADE

G. Kirby, B. Auchmann, M. Bajko, M.Charrondiere, N. Bourcey, V.I. Datskov, P. Fessia,
J. Feuvrier, P. Galbraith, A. Garcia Tabares, J. Garcia-Perez, P. Granieri, P. Hagen, C. Lorin,
J. C. Perez, S. Russenschuck, T. Sahner, M. Segreti, E. Todesco, G. Willering, CERN, Geneva,
Switzerland

*Abstract*

MQXC is a Nb-Ti quadrupole designed to meet the accelerator quality requirements needed for the phase-1 LHC upgrade, now superseded by the high luminosity upgrade foreseen in 2021. The 2-m-long model magnet was tested at room temperature and 1.9 K. The technology developed for this magnet is relevant for other magnets currently under development for the high-luminosity upgrade, namely D1 (at KEK) and the large aperture twin quadrupole Q4 (at CEA). In this paper we present MQXC test results, some of the specialized heat extraction features, spot heaters, temperature sensor mounting and voltage tap development for the special open cable insulation. We look at some problem solving with noisy signals, give an overview of electrical testing, look at how we calculate the coil resistance during at quench and show that the heaters are not working We describe the quench signals and its timing, the development of the quench heaters and give an explanation of an Excel quench calculation and its comparison including the good agreement with the MQXC test results. We propose an improvement to the magnet circuit design to reduce voltage to ground values by factor 2. The program is then used to predict quench Hot-Spot and Voltages values for the D1 dipole and the Q4 quadrupole.

## INTRODUCTION

For the phase-1 luminosity upgrade of the Large Hadron Collider at CERN, a development program was started in 2007 in collaboration with CEA-Saclay to develop a Nb-Ti 120 mm aperture quadrupole MQXC with an operational gradient of 120 T/m and the ability to extract heat loads of the order of 10 W/m. This quadrupole [1-4] had the innovative feature of an insulation scheme allowing a direct path from the helium bath to the superconducting strands [5].

After the decision of having only one upgrade, based on $Nb_3Sn$ technology for the inner triplet, the MQXC program is the backup plan for the upgrade. Moreover, it allows testing the novel insulation scheme that may be used in the upgrade for the separation dipoles D1 and D2, for the two-in-one quadrupole Q4, and for the orbit correctors [6, 7].

In this paper we describe the final assembly of the first 2-m-long model magnet, that was assembled at CERN; we also describe the test setup, as well as results for the training, quench performance, and quench location, magnet protection and in particular quench heaters efficiency. Special tests were carried out to study heat extraction, with encouraging results.

## NB-TI QUADRUPOLE FOR THE TRIPLET

### Magnet assembly

As for the LHC main dipole, the coil layers, made up of two different cables, are wound and cured to size individually [4]. Inner and outer layers are then assembled together with the quench heaters between the two coil layers. The coils are measured and the ends are shimmed so that the coil pressure gradually reduces (at room temperature) from the 80 MPa in the straight section to 30 MPa at the coil extremity. The four poles are sorted to optimize the coil mid-plane position. During the assembly, coils are placed vertically around a spring-loaded, collapsible mandrel and held in place with straps (see Fig. 1).

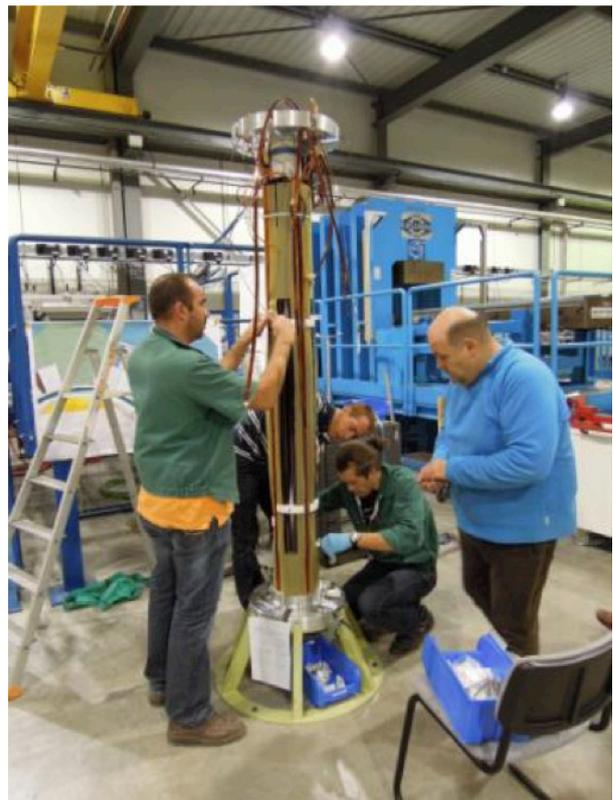

Figure 1: Vertical assembly of the coils around the collapsible mandrel, and open ground insulation around the coils.

The cooling sheets are mounted on the coils and pass through the ground insulation providing an open path to extract heat from the coil to the superfluid helium bath. Full-length heaters are placed between poles in order to simulate the beam-induced heat load. The full-length collaring shoes are then placed on top of the ground insulation to protect it from being damaged by the collars.

To further improve the magnet cooling, the collaring shoes are also perforated with openings of about 30% of the surface area (see Fig. 2). The 3-mm-thick Nippon stainless steel collars (with a ±0.01 mm tolerance) are stacked around the aperture and spaced to give a 3.3% open gap between the collars to extract heat. Eight holes in the collars, placed at 30° w.r.t. the mid-plane, can be filled with magnetic shims to optimize field quality. The aperture is locked with eight full-length keys using a collaring press. After this operation, the mandrel is removed. After welding the end flanges on to the collared aperture, the joints are soldered in the joint-box.

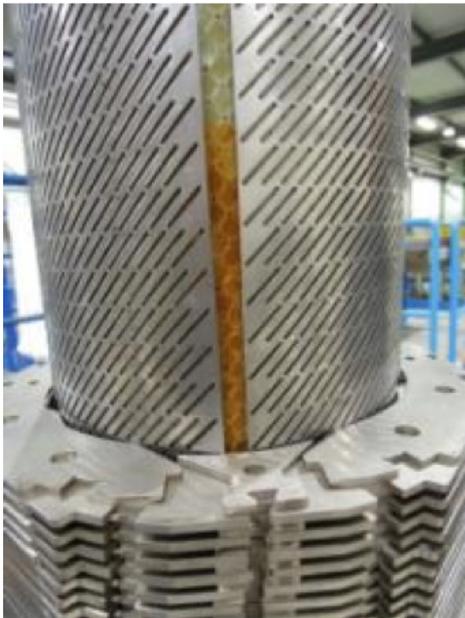

Figure 2: Assembly of the collared coil: view of collars before compression, and collaring shoes with openings.

The collared aperture is then placed vertically in the yoking tooling. The yoke laminations are stacked with an identical system as used for the LHC main quadrupole assembly. The obtained yoke packing factor has been 99.6%, i.e. larger than the expected 98%. The magnet is completed with the placement of the yoke end flange and mounting the four 80 mm diameter tie rods to provide longitudinal load.

During collaring, one of the magnet cables that exit a coil had three strands accidentally cut. Since this damage is in a low field region it was expected to only marginally affect the performance, hence it has been decided to continue without repairing the cable, which would have required a complete disassembly.

*Heat extraction features*

Principal features that contribute to the heat extraction are:
- The cable insulation;
- The open ground insulation;
- Perforated collaring shoes;
- The spacing between collars providing a 0.2 mm gap for helium at 1.9 K;
- Helium slots in quench heater to allow steady state heat extraction.

The open ground insulation is based on the idea of having a direct helium path through the insulating sheets to the strands, yet maintaining the voltage integrity by virtue of the voltage break down path length of ~ 20 mm. A plastic sheet 0.5-mm-thick with machined slots on both sides is used. This is placed on top of the coils on all surfaces that are in contact with the magnet structure, providing 30% film of helium over the full coil to extract heat. Then the layers of Kapton ground insulation start to be applied. A second 0.5-mm-thick sheet with the same machined slots is engineered to be in contact with the layer touching the coil and pass through the ground insulting sheets until it lies on top of the insulation yet under the perforated collaring shoes, see Fig 2. The machined sheets can just be seen under the perforated collaring shoes. The final collaring is done with a horizontal press (see Fig. 3).

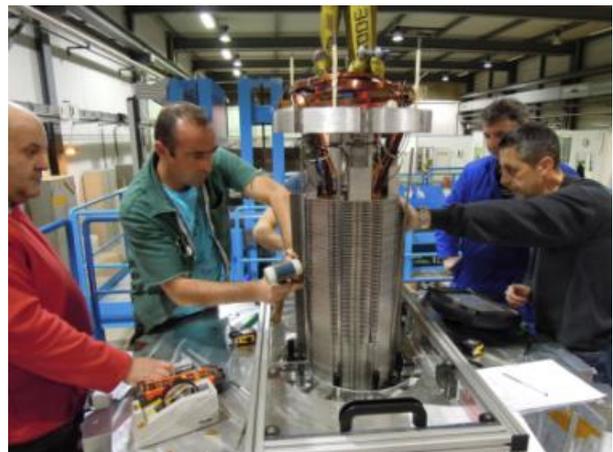

Figure 3: A view of the final stage of collaring.

The quench heaters are placed between the inner and the outer layer to act simultaneously on both layers; to improve the heat extraction during operation, quench heaters have slots to allow helium to flow from inner to outer layer, see Fig. 4. Moreover, during a quench heater firing, the helium in the slots would convert to gas and add to the acceleration of the quench over the uncovered coil surface. The quench heater was designed and tested to be hotter and faster than the LHC design. During test, we observed that the helium channels reduce the effect of the quench heaters, making them less efficient, so this design feature should be reviewed.

For the second model MQXC2 we have added two spot heaters, an array of voltage taps (see Fig. 5) and fast CCS temperature sensors to be able to measure the hot spot temperature in the coils (see Figs. 5, 6 and 7). We have evidence that using adiabatic assumption we significantly over estimate the hot spot temperature, due to the very efficient cooling through the cable, ground insulation, and open magnet structure.

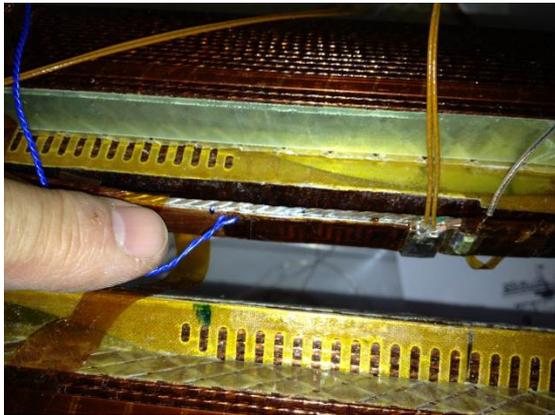

Figure 4: A view of the quench heaters with their cooling slots mounted between layers. We also see the spot heater and adjacent voltage taps.

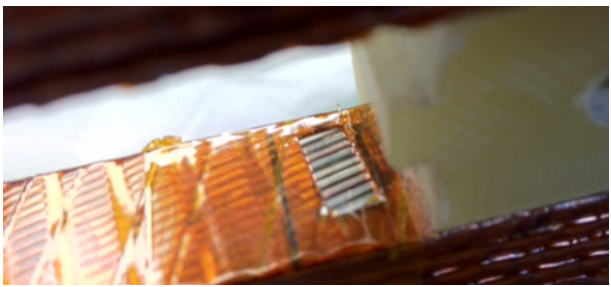

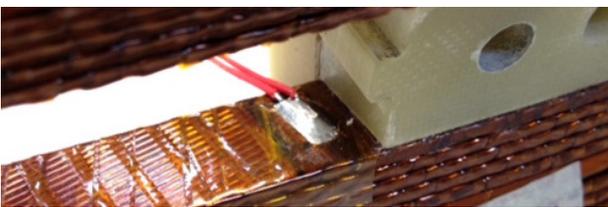

Figure 5: Window cut in cable insulation (upper part) to be able to mount temperature sensor or voltage taps on the cable (lower part).

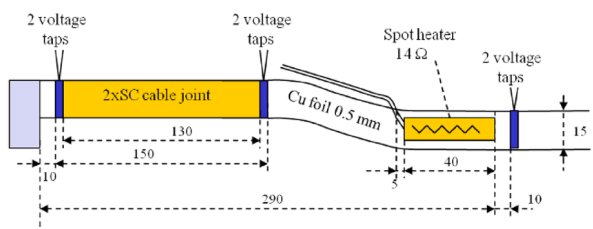

Figure 6: Schematic of spot heater position and voltage taps near the joint between inner and outer layer.

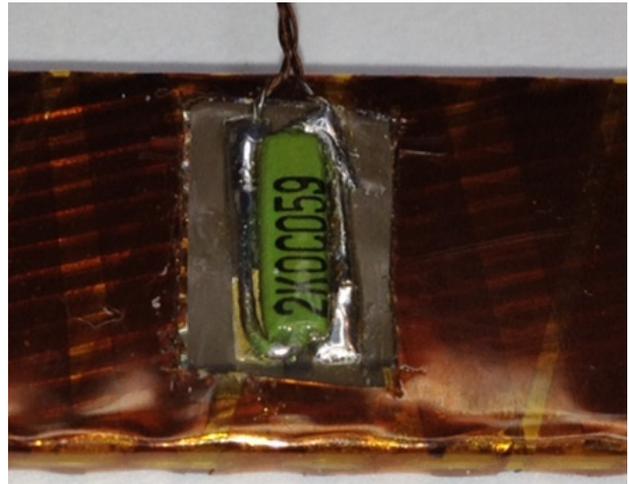

Figure 7: CCS temperature sensor mounted directly on cable. The sensor has a 0.025mm thick kapton film insulating it from the coil yet maintaining a rapid thermal response to temperature change during measurement.

The magnet joint resistance was measured during the powering tests and we found that there was a strong inductive element to the signal. After looking at photographs of how the voltage taps were routed out of the magnet we found a set of inductive loops, see Fig. 8. A correction will be implemented in the second model (see Fig. 9).

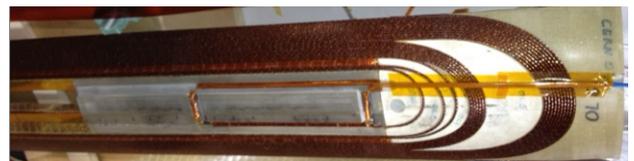

Figure 8: Inductive loop formed by the voltage taps coming from the interlayer joint in MQXC1.

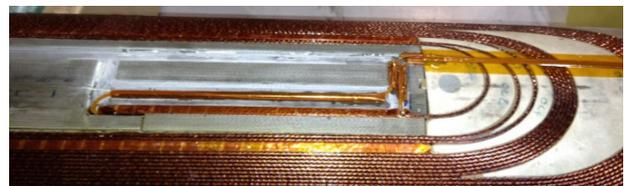

Figure 9: Correction to the loop adopted in MQXC2.

*Electrical tests*

The MQXC1 magnet also allowed to develop a comprehensive set of standard electrical test that will be applied to future magnets. The test starts with the coil after curing, still in the mould. The tests are repeated after each step of the magnet's assembly: coil winding, collaring, yoking, transport to test station, mounting on cold test support, and after insertion into the cryostat. A 1 kV pulse test looks for turn-to-turn shorts in the coil, using a resonating RCL circuit: a capacitor is charged to 1 kV and discharged into the coil. The four coils should have similar signals. If the exponential decay in a coil is slower than the others, the inductance is smaller and therefore one has lost turns due to an internal short. After collaring, the ground insulation is tested with 5 kV between the coils and ground. The main coil parameters are measured, i.e., resistance and inductance at a few different frequencies. The quench heaters are fired at room temperature with their full voltage.

Each voltage tap wire is connected in series with a 10 kΩ resistor. This resistor protects the wiring in the event of an electrical short. All the instrumentation wiring from the magnet exits through the lambda plate feed-thoughts and out of the cryostat. The analog signals travel approximately 20 m from the cryostat to the analog-to-digital cards, where the quench trigger thresholds are set.

*Magnet circuit*

The magnet and the circuit during test are shown in Fig. 10. The 20 kA power converter is grounded on the negative side of the converter. Later we will discuss an improved position for the grounding point. The converter only has positive voltage so the negative ramps are driven by the decay through the room temperature current lead resistance.

The protection switch and dump resistor are large components. The dump can be configured to give combinations of the 4 × 20 mΩ resistors connected in series or parallel or combinations.

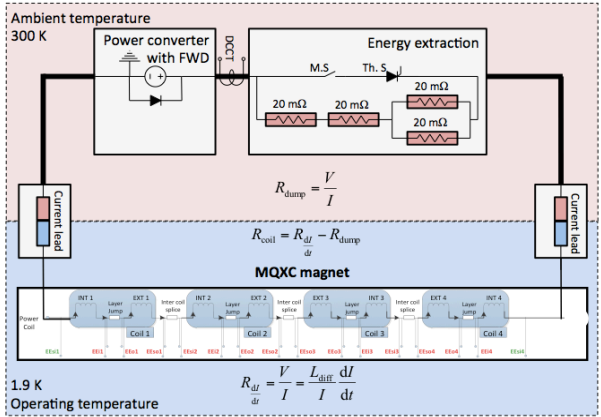

Figure 10: The magnet circuit in test stand.

*Differential inductance measurements*

The inductance measurement is performed by ramping up and down from 80 A to 12800 A at the nominal ramp-rate of 11 A/s. The inductance was deduced for the inner and outer layers of coil 1 to 3 separately, for the inner and outer layer of coil 4 combined and for the full magnet, see results in Fig. 11 and 12.

We see a significant hysteresis between 80 A and 2 kA due to the magnetization of the filaments. Estimates through a ROXIE [10] model are in good agreement with the measurements for the full magnet and show that the coil inductance is constant up to about 5 kA. Above 5 kA the inductance decreases due to saturation in the magnetic yoke.

When summing the inductance of the separate parts, the values do not add up to the full magnet inductance, because the voltage taps were wrongly installed forming a pickup coil and hence reducing the measured inductance. Additionally, inductances of 0.07 to 0.17 mH were measured with the voltage taps across inter-layer joints and inter-coil joints, which also indicate that voltages were picked up. In a next measurement the voltage taps will be changed such that the pick-up voltage is minimised.

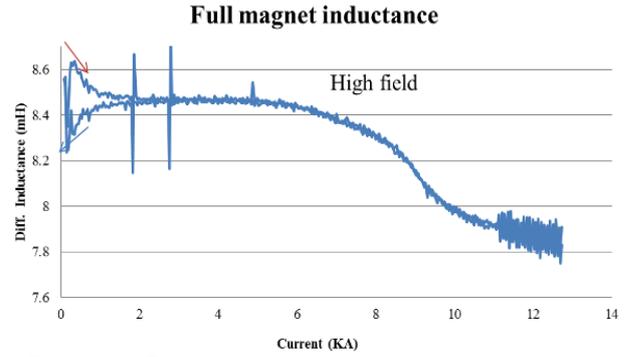

Figure 11: Differential inductance measured at 11 A/s.

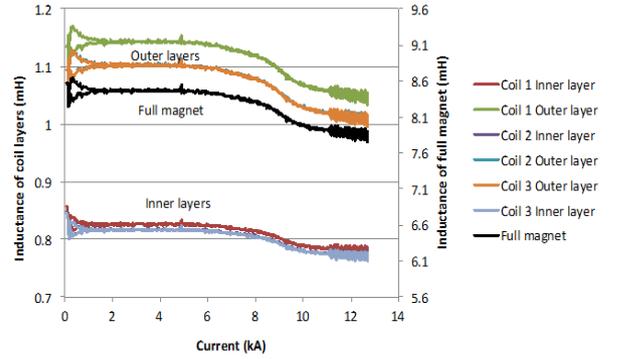

Figure 12: Differential inductance measured in each coil (left scale), and in the full magnet (right scale).

*Quench-back test*

A special test to study the quench-back has been done: the idea is ramping to nominal, open the switch and disconnecting the power supply, dumping the current into the external resistor. Then the instantaneous resistance of the coil $R_c(t)$ is estimated from the derivative of the current w.r.t. time

$$\frac{dI(t)}{dt} = -\frac{I(t)}{\tau(t)} \qquad \tau(t) = \frac{L(t)}{R_c(t) + R_d}$$

Where the nonlinearity of the inductance is taken into account, i.e. at each instant $t$ we use the inductance $L(I(t))$. In a similar test, the $Nb_3Sn$ quadrupole HQ developed a significant resistance (i.e. it quenched) due to the fast initial ramp rate [9]. In our case we see a very limited development of resistance of about 10 mΩ, see Fig. 13.

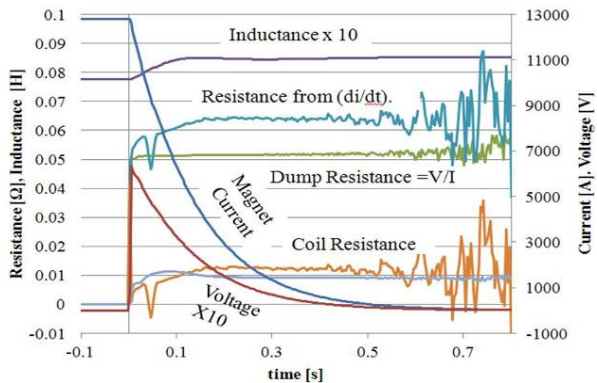

Figure 13: Resistance versus time during the dump of the current on external resistor at nominal current.

## Quench heater performance

In Fig. 14 we estimated the resistance needed to protect the magnet as a function of the magnet current. We assumed a (fast) detection time of 12 ms. The light-green plot gives the circuit resistance needed to limit the hot spot to 100 K, a very conservative value where the coil thermal expansion is extremely low and will not induce any mechanical movement. At the operational value of 12.8 kA, 50 mΩ are needed. The purple plot is the circuit resistance needed to limit the hot spot temperature to 300 K. This is the limit we assume to avoid degradation: 20 mΩ are needed at operational current.

In the same figure, the red line is the measured coil resistance developed during test quenches. The detail of these measurements is shown in Fig. 15: the magnet quenchback does not provide significant resistance. The magnet was tested with the 50 mΩ dump resistance so it was not harmed. However if the dump was not activated, at nominal current 12800 A the adiabatic hot spot temperature is predicted to be 1200 K. So the conclusion is that the internal coil resistance is insufficient to protect the magnet.

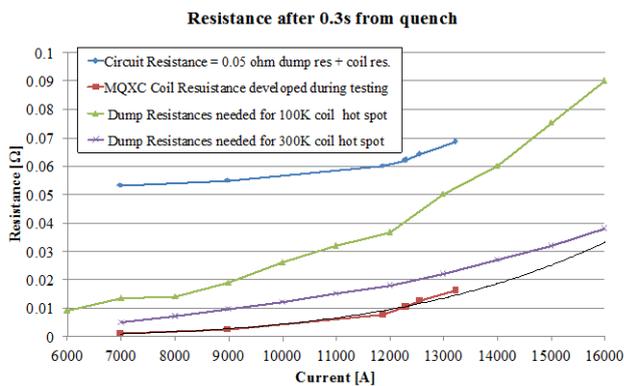

Figure 14: Circuit resistance needed to protect the MQXC model magnets as a function of magnet current.

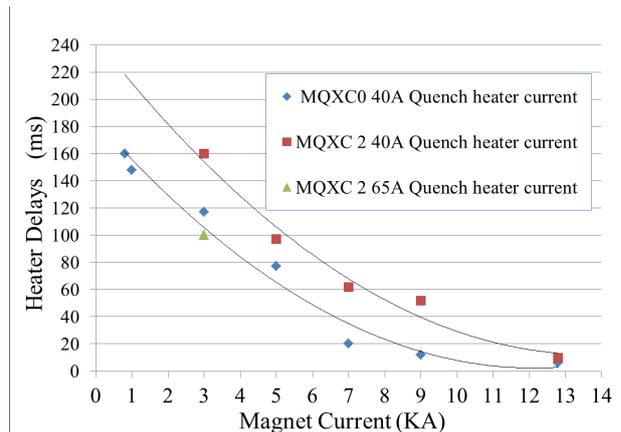

Figure 15: Quench heater delays with 40 A and 80 A in the heaters (markers) and parabolic fit.

In Fig. 15 we show the delay of the quench induced by heaters as a function of the current during a quench triggered by the heaters. At nominal current one has values of the order of 10 ms or less.

We checked not to overcome a temperature of 200 K in the quench heaters. We used a 200 Hz infrared camera to read the temperature after discharge at room temperature (see Fig. 16). Thermocouples mounted on the heater (see Fig. 17) allowed to measure the temperature in operational conditions, and to distinguish between copper plated and stainless steel zones (see Fig. 18).

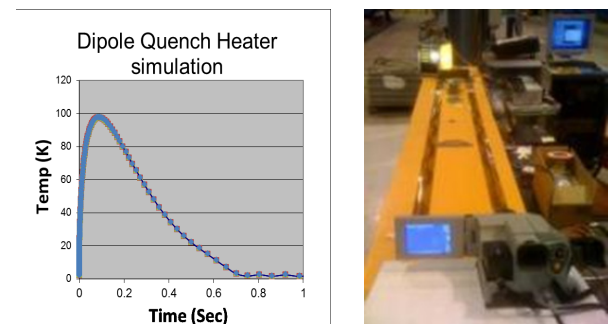

Figure 16: Calculated heater response and view of infer red camera measurement at room temperature.

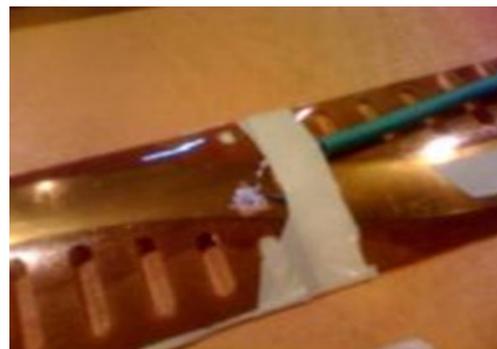

Figure 17: Thermocouple mounted directly on stainless steel heater element, with the Kapton insulation cut away.

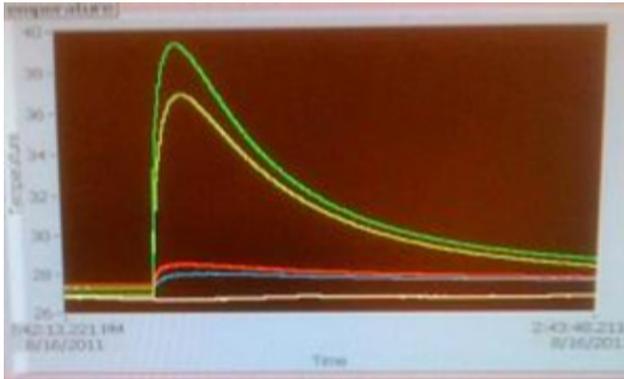

Figure 18: Thermocouple reading on stainless steel (green and yellow) and copper plated zones (red and blue). Reference thermocouple is in white.

*Finite difference model*

The quench calculation for assessing the magnet performance used a finite difference approach, implemented in an excel spread sheet, including the temperature dependence of the specific heat of the cable and the copper resistivity. The modelled circuit includes differential inductance for the magnet, quench heater delays for different parts of the coil, energy extraction to the resistive dump and resistance of the room temperature current leads. Althow we see good agreement with the calculated currents and voltages, we have no verification with a measurement of the hot spot temperature. An example of test results and the model results are given in Fig. 19 and 20.

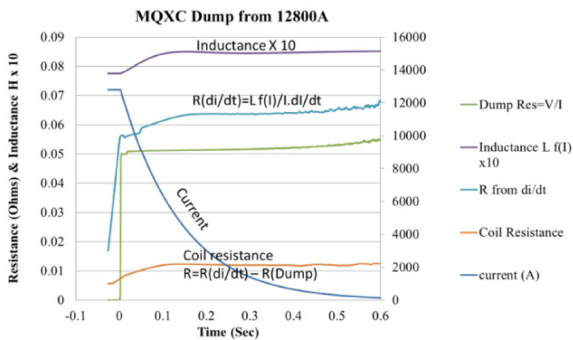

Figure 19: Test results for a quench at 12.8 kA.

*Full length MQXC protection circuit*

As the quench heaters still need development, one simple effective solution and safe alternative is to power the four insertion triplet magnets individually. The quench simulation for the 9.5-m-long magnet was performed. With an external energy extraction, a dump resistor of 130 mΩ, and 0.016 s delay this gives a hot spot of just over 250 K and the main bulk of the magnet is about 100 K. This uses the idea of placing the earth at the centre of the dump resistor, allowing to double the dump resistor value without increasing the voltage. Turn-to-turn and layer-to-layer voltages are unchanged between the earth configuration positions.

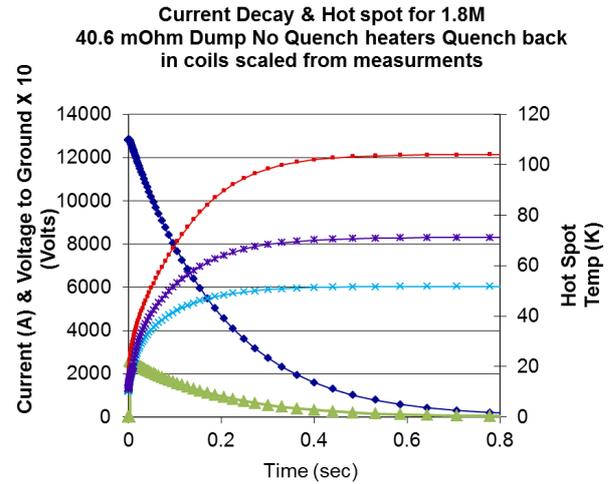

Figure 20: Model results for quench at 12.8 kA. Red is hot spot, light blue is outer layer, and purple is inner layer average coil block layer temperatures, Green is voltage across dump times 10, dark blue box is the current decay.

## SEPARATION DIPOLE

For the separation dipole D1, single aperture, with an operational field of ~5 T and a total length of ~7 m, KEK colleagues are considering to using the LHC main dipole outer cable, and possibly the insulation used on MQXC, to take advantage of the high heat extraction that may be needed for this magnet.

The first quench study looked to see if the magnet could be protected with quench heaters as is standard in LHC large magnets. The 7-m-long magnet has a large bore, so the inductance to resistive coil ratio is high. The study showed that without dump we would need to quench 100% of the coil in 0.016 sec, reaching 260 K (see Fig. 21). On the other hand, with a 100 mΩ dump resistor and the same delay the hotspot temperature is below 200 K (see Fig. 22).

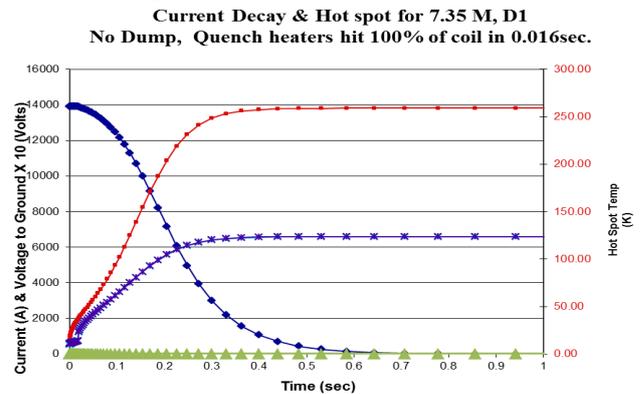

Figure 21: D1 quench simulation without dump resistor and with unrealistically fast heaters and quench coverage. Red curve is the hot spot, blue square is the current decay, purple is the average coil temperature, and green is the dump voltage.

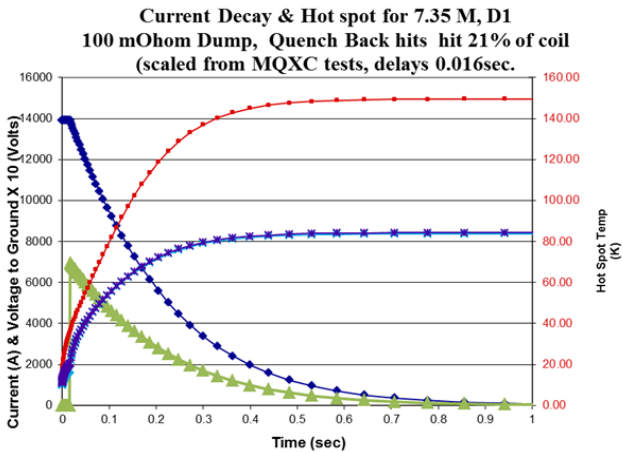

Figure 22: Proposed protection scheme with 100 mΩ dump resistor and quench heaters. Red curve is the hot spot temperature, purple square is the current decay, blue marker is the average coil temperature, and green is the dump voltage times 10.

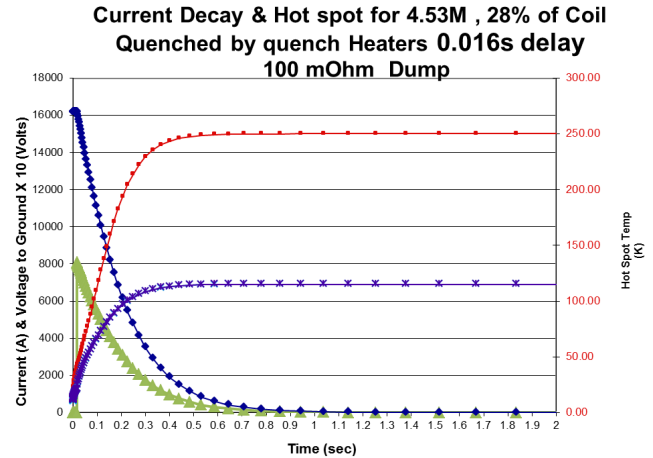

Figure 24: Proposed protection scheme with 100 mΩ dump resistor and quench heaters. Red curve is the hot spot, purple square is the current decay, purmpe X in the average coil temperature, green in dump voltage x10.

## LARGE APERTURE TWO-IN-ONE QUADRUPOLE

The Q4 under design at CEA (see Fig. 23) has a large aperture and could also possibly use the LHC dipole cable with the enhanced cable insulation as developed for MQXC. The heat load for this magnet can be high, so also in this case the cooling will be important. All the data for this magnet and others can be found at: www.cern.ch/hilumi/wp3. As for D1, energy extraction dump resistors over each aperture of the 4.5-m-long magnet limit the maximum hot spot temperature and maximum voltage to ground to an estimated 250 K and 800 V, respectively (see Fig. 24). The average temperature of the coil is at ~115 K. for this configuration.

## CONCLUSIONS

MQXC is the Nb-Ti option for the High-Luminosity LHC upgrade. It has been designed to maximize the cooling leaving open paths for HeII to the strands. Test results have shown some conflict between the need of a large heat extraction and the needs of quench protection. With a dump resistor the magnets proves to be protectable. Instrumentation has been installed for the next round of magnet tests to better understand hot spot and quench properties. Quench studies for both D1 and Q4 magnets, foreseen for the HL-LHC, indicate that a dump resistor can guarantee a safe protection scheme.

## REFERENCES

[1] J. P. Kouthchouk, L. Rossi, E. Todesco, LHC project Report 1000 (2006).
[2] R. Ostojic, et al., Conceptual design of the LHC interaction upgrade, LHC Project Report, 1163 (2008).
[3] S. Russenschuck et al., Design challenges for a wide aperture superconducting quadrupole, IEEE Transactions on Applied Superconductivity, 21 (2011), 1674-7.
[4] G.A.Kirby, et al. Engineering Design and Manufacturing Challenges for a Wide-Aperture, Superconducting Quadrupole Magnet, IEEE Transactions on Applied Superconductivity, 22 (2012).
[5] M. La China, D. Tommasini, Cable insulation scheme to improve heat transfer to superfluid helium in Nb-Ti accelerator magnets, IEEE Transactions on Applied Superconductivity, 18 (2008), 1285.
[6] L. Rossi, these proceedings.
[7] Q. Xu, these proceedings.
[8] J. García Pérez, J. Billan, M. Buzio, P. Galbraith, D. Giloteaux, V. Remondino, Performance of the Room Temperature Systems for Magnetic Field

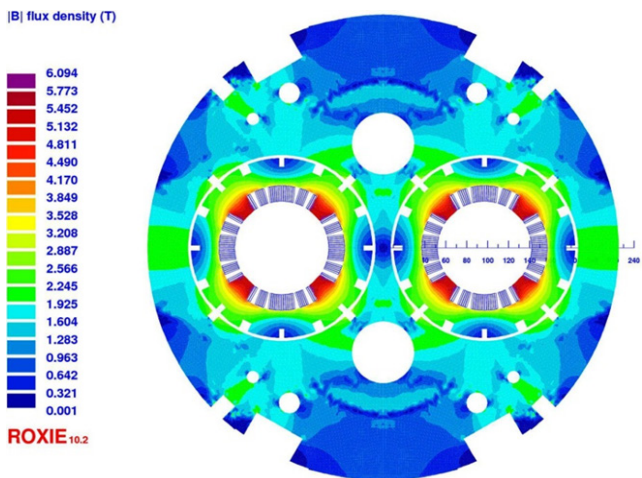

Figure 23: Q4 cross section